\documentclass[12pt,a4paper]{article}
\begin{document}

\title{ Klein's Paradox } 
\author {A. Bounames \\
 \it{Laboratoy of Theoretical Physics, Faculty of Science,}\\ 
 \it {University of Jijel, BP 98 Ouled Aissa, 18000  Jijel, Algeria}  
\and L. Chetouani \\  
\it{D\'epartement de Physique, Facult\'e des Sciences Exactes, Universit\'e} \\ \it{Mentouri, Route Ain El-Bey, 25000  Constantine, Algeria}}

\date{}
\maketitle

\vspace{0.5cm}

\maketitle

\begin{abstract}
 
 We solve the one dimensional Feshbach-Villars equation for spin-1/2 particle
subjected to a scalar smooth potential. The eight component wave function is 
given in terms of the hypergeometric functions and via a limiting procedure,
the wave functions of the step potential are deduced. These wave functions
are used to test the validity of the boundary conditions deduced from the 
Feshbach-Villars transformation. The creation of pairs
is predicted from the boundary condition of the charge density.\\

\end{abstract}


\newpage
   The eight component relativistic wave equation for spin-1/2 particle, 
called the Feshbach-Villars equation $(FV\frac{1}{2})$, has been constructed
and used to solve physical problems \cite{rob1,rob2,stau1,stau2}. The hydrogen atom is the 
first problem solved by Robson and Staudte \cite{rob1,rob2}; they found the
same bound-state energy as the Dirac equation but the wave functions are
different. Recently, Robson and Sutanto have solved the Compton  scattering
problem and found that the cross section is given, like in Dirac theory,  by
the Klein-Nishina formula \cite{rob3}. The same authors have also calculated 
the transition probabilities  for the Balmer and Lyman $\alpha$-lines of
hydrogenic  atoms and compared them to the Dirac and Schrodinger results \cite{rob4}.

Following the idea of a previous paper \cite{mer},
we study the one dimensional $FV\frac{1}{2}$ equation for
a particle subjected to a step potential $V(x)$. As we know the above problem 
is trivial in quantum mechanics but in relativistic quantum mechanics it 
emerges the famous Klein paradox. In addition to the known literature in 
this framework \cite{domb1}, we limit ourselves in this paper to illustrate the problem more clearly and justify the phenomenon of pair creation, and we left the problem of 
interpretation to the specialist researchers.
The boundary conditions for the eight component wave function 
for the case of the step potentiel are unknown. Then, in order to bypass
this problem, we take $V(x)$ as a smooth potential \cite{che1,che2,che3}. The analytic solution of 
the FV-1/2 equation with the smooth potential is given. We deduce, via a limiting procedure, the wave functions of the step potential and derive the transmission and reflection coefficients. 
The comparison with the Dirac coefficients constitutes the test to our calculations.
The appropriate boundary conditions for the step potential are extracted from the 
Feshbach-Villars transformation. The boundary condition  for 
the charge density is also evaluated. The validity of these boundary conditions 
is tested using the wave functions of the step potential. 
At the end, we discuss the boundary condition of the charge density 
and compare its predictions to those obtained from the transmission and reflection coefficients.

The Dirac equation can be written in a second-order form as 
\begin{equation}
\bigl[ (\gamma ^{\mu}D_{\mu})^2+m^2 \bigr] \Phi= 0, 
\end{equation}
where $\gamma^{\mu}$ are the Dirac matrices, $D_{\mu}$ is the minimally coupled 
derivative and $\Phi({\varphi_{1},\varphi_{2},\varphi_{3},\varphi_{4}})$
is the four component wave function.
This equation can be also written in the Klein-Gordon form as follows \cite{fey}
\begin{equation}
\bigl[ (D_{\mu} D^{\mu}+m^2){\bf 1_4}+\frac{e}{2}\sigma^{\mu\nu} F_{\mu\nu}
\bigr]\Phi= 0, \label{kg}
\end{equation}
where $\sigma^{\mu\nu}=\frac{i}{2}\bigl[\gamma^{\mu},\gamma^{\nu} \bigr]$ and 
$F_{\mu\nu}=\partial_{\mu}A_{\nu}-\partial_{\nu}A_{\mu}$. 
The last term in Eq. (\ref{kg}) represents the spin
interaction with the external electromagnetic field $F_{\mu\nu}$.  By analogy
with the equation of spin-0 particle, the second order equation (\ref{kg}) is
called the Klein-Gordon equation for spin-1/2 particle
$(KG\frac{1}{2})$ \cite{rob1,rob2}. For spin-0 particles, this equation  is
reduced to a Klein-Gordon type equation.

In order to linearize Eq. (\ref{kg}) to a first order equation in time, the Feshbach-Villars linearization procedure is used to transform the four component wave function $\Phi$ to an eight 
component wave function $\psi$ \cite{rob1,rob2}. The eight  component wave 
function $\psi$ satisfies a Schrodinger type equation \cite{rob2}
\begin{equation}
H\psi =i \frac{\partial }{\partial t} \bf{1}_8 \psi,  \label{fv}  
\end{equation}
with 
\begin{equation}
H=(\tau_{3}+i\tau_{2})\otimes\frac{1}{2m}\left[-{\bf D^2 1_4 }+\frac{e}{2}
\sigma^{\mu\nu} F_{\mu\nu}\right]+m(\tau_{3}\otimes{\bf 1_4})+eA_{0}{\bf 1_8},      
\end{equation}
where $\tau_{2}$, $\tau_{3}$ are the Pauli matrices, $\otimes$ is the
Kroneker(direct) product, ${\mathbf D}={\mathbf \partial}+ie{\mathbf A}$  is tridimensional minimally coupling and
$({\mathbf A},A_{0})$ is the electromagnetic potential.

The Hamiltonian $H$ is pseudo-Hermitian $H=\tau _{4}H^{\dagger }\ \tau_{4}$ 
and the inner product is 
\begin{eqnarray}
\left( \psi ,\psi \right) =\int \psi^{\dagger} \tau_4 \psi d^{3}V, \qquad 
 \label{sca}  \nonumber
\end{eqnarray}
where $\tau_4=\tau_3 \otimes \gamma^0$. This inner product is indefinite in sign, it takes positive or negative values. Then, the (FV-1/2) wave function space is not a Hilbert space and its dimension is twice that of the Dirac solution space \cite{rob2}.

In the Weyl representation of the gamma matrices, Eq. (\ref{fv}) separates into two four-component equations with the Hamiltonians
\begin{equation}
H_{\xi}=(\tau_{3}+i\tau_{2})\otimes\frac{1}{2m}\left[- {\bf D^2}
{\bf 1_2}+ie{\mathbf \sigma .}({\mathbf E}+i {\mathbf B})\right]
+m(\tau_{3}\otimes {\bf 1_2})+eA_{0}{\bf 1_4},  \label{ham1}
\end{equation}
\begin{equation}
H_{\eta}=(\tau_{3}+i\tau_{2})\otimes\frac{1}{2m}\left[-{\bf D^2}
{\bf 1_2}-ie{\mathbf \sigma .}({\mathbf E}-i{\mathbf B})\right]
+m(\tau_{3}\otimes {\bf1_2})+eA_{0}\bf{1_4}, \label{ham2}
\end{equation}
where $\mathbf{E}$, $\mathbf{B}$ are the intensities of the electromagnetic
field and ${\mathbf \sigma}(\tau_{1},\tau_{2},\tau_{3})$ are the Pauli
matrices.
 
The Hamiltonians $H_{\xi}$ and $H_{\eta}$ satisfy also a
Schrodinger type equation 
\begin{equation}
H_{\xi}\psi_{\xi} =i \frac{\partial }{\partial t} {\bf 1_4} \psi_{\xi}, \label{sch1} 
\end{equation}
\begin{equation}
H_{\eta}\psi_{\eta} =i \frac{\partial }{\partial t} {\bf 1_4} \psi_{\eta}, \label{sch2} 
\end{equation} 
where $\psi_{\xi}$ and $\psi_{\eta}$ are four-component wave functions
defined by their components as 
$$\psi_{\xi}= \left(\psi_{1},\psi_{2},\psi_{3},\psi_{4}\right)^T,\qquad
\psi_{\eta}=\left(\psi_{5},\psi_{6}, \psi_{7},\psi_{8}\right)^T,$$
and the eight component wave function is $\psi=\left(\psi_{\xi},\psi_{\eta}\right)^T$.  The Hamiltonians $H_{\xi}$ and $H_{\eta}$, the wave functions $\psi_{\xi}$ and
$\psi_{\eta}$ transform to each other under spatial
inversion, respectively \cite{rob1,rob2}.

In the Weyl representation of the gamma matrices, the density $\rho $ is defined \cite{rob2} as follows 
\begin{eqnarray*}
\rho =\overline{\psi}\psi  \quad   \mbox{ where }  \quad   \overline{\psi}={\psi^{\dag}}\tau_{5}, 
\quad  \tau_{5}=\tau_{1}\otimes(\tau_{3}\otimes {\mathbf 1_{2}}),
\nonumber
\end{eqnarray*}
where $\rho$ and $\tau_{5}$ are given by Eqs. (62),(63) in Ref. \cite{rob2}.\\
We define the one dimensional current $j$ by  
\begin{eqnarray*}
j=\frac{1}{2im}\left[ \overline{\psi}O \frac{\partial \psi }{\partial x}-\frac{\partial \overline{\psi}}{\partial x} O \psi \right] 
-\frac{e}{m}A \overline{\psi } O \psi,\quad \mbox{ where } \quad  
O={\mathbf 1_{2}}\otimes(\tau _{3}+i\tau _{2})\otimes {\mathbf 1_{2}}.
\nonumber 
\end{eqnarray*}

The values of $\rho$ and $j$ are independent of the representation and they satisfy the continuity equation.  $\rho $ is interpreted as the charge density of the particle. The positive
solution $\psi$, the negative solution $\psi _{c}$, the charge density and
the current are transformed by the charge conjugation as follows
$$\psi \longrightarrow \psi _{c}=\tau _{1}\otimes {\gamma^{0}} \psi ^{\dagger }, \quad \rho
\longrightarrow \rho _{c}=-\rho \qquad \mbox{ and } \quad  j\longrightarrow
j_{c}= j.$$  

In one dimension, Hamiltonians (\ref{ham1}),(\ref{ham2}) for
spin-1/2 particle subjected to a scalar potential $V(x)$ are
\begin{equation}
H_{\xi}=(\tau_{3}+i\tau_{2})\otimes\frac{1}{2m}\left[- \frac{d^{2}}{dx^{2}}
{\bf 1_2}-ie\tau_1 \frac{dV(x)}{dx} \right]+m(\tau_{3}\otimes
{\bf 1_2})+eV(x){\bf 1_4}, \label{hd1} 
\end{equation}
\begin{equation}
H_{\eta}=(\tau_{3}+i\tau_{2})\otimes\frac{1}{2m}\left[-\frac{d^{2}}{dx^{2}}
{\bf 1_2}+ie \tau_1 \frac{dV(x)}{dx} \right]+m(\tau_{3}\otimes
{\bf1_2})+eV(x)\bf{1_4}. \label{hd2}
\end{equation}
The terms $\pm ie\tau_{1}\frac{dV(x)}{dx}$ represent the interaction of the
spin with the derivative of the scalar potential.

The stationary solution has the form 
$\psi(x,t)=e^{-iEt}\psi(x)$ for each component of the wave function and Eq. (\ref{sch1}) 
of the Hamiltonian $H_{\xi}$ is equivalent to the
following four coupled differential equations 
\begin{equation}
\left[-\frac{d^{2}}{dx^{2}}+2m^2+2meV(x)-2mE\right]\psi_{1}-\frac{d^{2}\psi_{3}}{dx^{2}}
-ie\frac{dV(x)}{dx}(\psi _{2}+\psi _{4})=0, \label{eq1}
\end{equation}
\begin{equation}
\left[-\frac{d^{2}}{dx^{2}}+2m^2+2meV(x)-2mE\right]\psi_{2}-\frac{d^{2}\psi_{4}}{dx^{2}}
-ie\frac{dV(x)}{dx}(\psi _{1}+\psi _{3})=0, \label{eq2}
\end{equation}
\begin{equation}
-\frac{d^{2}\psi_{1}}{dx^{2}}+\left[-\frac{d^{2}}{dx^{2}}+2m^2-2meV(x)+2mE\right]\psi_{3}
-ie\frac{dV(x)}{dx}(\psi _{2}+\psi _{4})=0, \label{eq3}
 \end{equation}
\begin{equation}
-\frac{d^{2}\psi_{2}}{dx^{2}}+\left[-\frac{d^{2}}{dx^{2}}+2m^2-2meV(x)+2mE\right]\psi_{4}
-ie\frac{dV(x)}{dx}(\psi _{1}+\psi _{3})=0. \label{eq4}
\end{equation}
The difference of Eqs. (\ref{eq1})-(\ref{eq3}) and (\ref{eq2})-(\ref{eq4})
give, respectively,   
\begin{equation}
\psi_{1}-\psi_{3}=\left[ \frac{E-eV(x)}{m}\right] (\psi_{1}+\psi_{3}),\label{dif1}  
\end{equation}
\begin{equation}
\psi_{2}-\psi_{4}=\left[ \frac{E-eV(x)}{m}\right] (\psi_{2}+\psi_{4}).\label{dif2}  
\end{equation} 
Using these equations, the sum of Eqs. (\ref{eq1})-(\ref{eq3})
and (\ref{eq2})-(\ref{eq4}) give   
\begin{equation}
\left[\frac{d^{2}}{dx^{2}}+\left[ E-eV(x)\right] ^{2}-m^{2}%
\right](\psi_{1}+\psi_{3})+ie\frac{dV(x)}{dx}(\psi_{2}+\psi_{4})=0, \label{e1}
\end{equation}
\begin{equation}
\left[\frac{d^{2}}{dx^{2}}+ \left[ E-eV(x)\right] ^{2}-m^{2}%
 \right](\psi_{2}+\psi_{4})+ie\frac{dV(x)}{dx}(\psi_{1}+\psi_{3})=0. \label{e2}
\end{equation}
 The sum and the difference of the two last equations (\ref{e1}),(\ref{e2}) give
\begin{equation}
\left[\frac{d^{2}}{dx^{2}}+\left[ E-eV(x)\right] ^{2}-m^{2}%
+ie\frac{dV(x)}{dx}\right]\psi_{\xi}^{s}(x)=0, \label{e3}
\end{equation}
\begin{equation}
\left[\frac{d^{2}}{dx^{2}}+\left[ E-eV(x)\right] ^{2}-m^{2}%
-ie\frac{dV(x)}{dx}\right]\psi_{\xi}^{d}(x)=0, \label{e4}
\end{equation}
where
\begin{equation}
\psi_{\xi}^{s}(x) = \psi _{1}+\psi _{2}+\psi_{3}+\psi_{4}, \qquad 
\psi_{\xi}^{d}(x)= \psi_{1}+\psi_{3}-\psi _{2}-\psi_{4}.  \label{xi}
\end{equation}
We note that the differential equation (\ref{e3}) is the same as the Dirac equation gives
(Eq.(207.7) in Ref. \cite{flug}).
If we take $V(x)$ as a step potential $V(x)=V_{0} \theta(x)$, we
have in Eqs. (\ref{e3}),(\ref{e4}) the potential $V(x)$ and its 
derivative: $V^{'}(x)=\frac{dV(x)}{dx}=V_{0}\delta(x)$. 
In the configuration space, the delta Dirac potential 
has particular treatment of its boundary conditions \cite{roy,ben}. In this paper, our goal is to
find the wave functions  without using boundary conditions. At the end, we use
the obtained wave functions to test the validity of the boundary conditions
deduced from the Feshbach-Villars transformation. For this, we use the 
scalar smooth potential $V(x)$ defined as  
\begin{equation}
V(x)=\frac{V_{0}}{2}(1+\tanh \frac{x}{2r}),
\end{equation}
where $V_{0}$ and $r$ are positive constants.
In the limiting case $r\rightarrow 0$, $\ V(x)\rightarrow V_{0}\theta (x)$.
It increases from the value $V=0$ \ for $x=-\infty $ \ to the value $V=V_{0}$
\ for $x=+\infty $, the main rise occurring in the interval \ $-2r<x<+2r:$
$V(-2r)=0.1192V_{0},$  $V_{0}(2r)=0.8807V_{0}.$

In order to find the solution of the differential equations (\ref{e3}),(\ref{e4}), 
we make the change of variable \begin{equation}
y=\frac{1}{2}(1-\tanh \frac{x}{2r}),
\end{equation}
which maps the interval $x\in \left] -\infty ,+\infty \right[ $ \ to $\ y\in %
\left] 0,1\right[ $ . The new form of Eqs. (\ref{e3}), (\ref{e4}) are
\begin{eqnarray}
\frac{1}{r^{2}}y^{2}(1-y)^{2}\frac{d^{2}\psi_{\xi}^{s}(y)}{dy^{2}} &+&
\frac{1}{r^{2}}y(1-y)(1-2y)\frac{d\psi_{\xi}^{s}(y)}{dy} 
\nonumber \\ 
&+&\left[(E+eV_{0}y-eV_{0})^{2}-m^{2}+i\frac{eV_{0}}{r}y(1-y)\right]\psi_{\xi}^{s}(y)=0, 
\label{frac1}
\end{eqnarray}
\begin{eqnarray}
\frac{1}{r^{2}}y^{2}(1-y)^{2}\frac{d^{2}\psi_{\xi}^{d}(y)}{dy^{2}} &+&
\frac{1}{r^{2}}y(1-y)(1-2y)\frac{d\psi_{\xi}^{d}(y)}{dy} 
 \nonumber \\
 &+&\left[(E+eV_{0}y-eV_{0})^{2}-m^{2}-i\frac{eV_{0}}{r}y(1-y)\right]\psi_{\xi}^{d}(y)=0. 
\label{frac2}
\end{eqnarray}
The singularities of these differential equations are $y=0,1,\ \infty .$ Let
us introduce the change $\psi _{\xi}^{s}(y)= y^{\nu}(1-y)^{\mu}{}f(y),$ and
$\psi _{\xi}^{d}(y)= y^{\nu }(1-y)^{\mu }{}g(y),$  the last
equations are reduced to the hypergeometric equation form
\begin{equation}
y(1-y)\frac{d^{2}f(y)}{dy^{2}}+\left[ (2\nu +1)-y(2\nu +2\mu +2)\right] 
\frac{df(y)}{dy}-\left[(\mu +\nu
+\frac{1}{2})^2-\frac{v_{1}^2}{4}\right]f(y)=0. \label{hyp1} \end{equation}
\begin{equation}
y(1-y)\frac{d^{2} g(y)}{dy^{2}}+\left[ (2\nu +1)-y(2\nu +2\mu +2)\right] 
\frac{dg(y)}{dy}-\left[(\mu +\nu+\frac{1}{2})^2
-\frac{v_{2}^2}{4}\right]g(y)=0, \label{hyp2}   
\end{equation}
where $ \nu ^{2}=r^{2}\left[ m^{2}-(E-eV_{0})^{2}\right], \quad  \mu^{2}=r^{2}(m^{2}-E^{2}),  
\quad  v_{1}^2=1-4r^{2}e^{2}V_{0}^{2}+4ireV_{0} $  \quad and\\
$ v_{2}^2=1-4r^{2}e^{2}V_{0}^{2}-4ireV_{0}.$

In comparison with the spin-0 case, the imaginary part in the expressions of $v_{1}^2$ and $v_{2}^2$
represent the effect of the spin. If we remove them, Eqs. (\ref{hyp1}), (\ref{hyp2}) are reduced to the same equation as for the spin-0 case \cite{mer}.

The general solutions of Eqs. (\ref{hyp1}),(\ref{hyp2}) are given in terms of the hypergeometric function
\begin{eqnarray*}
\psi_{\xi}^{s}(y) &=& C_{1}\ y^{\nu }(1-y)^{\mu }{}\ _{2}F_{1}(\mu +\nu
+\frac{1}{2}-\frac{v_{1}}{2},\mu +\nu+\frac{1}{2}+\frac{v_{1}}{2},1+2\nu,y) \\ &&
+C_{12}\ y^{-\nu }(1-y)^{\mu }{}\ _{2}F_{1}(\mu-\nu+\frac{1}{2}+\frac{v_{1}}{2}, \mu
-\nu+\frac{1}{2}-\frac{v_{1}}{2},1-2\nu,y),  \\
\psi_{\xi}^{d}(y) &=& D_{1}\ y^{\nu }(1-y)^{\mu }{}\ _{2}F_{1}(\mu +\nu
+\frac{1}{2}+\frac{v_{2}}{2},\mu +\nu +\frac{1}{2}-\frac{v_{2}}{2},1+2\nu,y) \\ &&
+D_{12}\ y^{-\nu }(1-y)^{\mu }{}\ _{2}F_{1}(\mu -\nu+\frac{1}{2}-\frac{v_{2}}{2},
\mu -\nu+\frac{1}{2}+\frac{v_{2}}{2},1-2\nu,y), 
\end{eqnarray*}
where $C_{1}$, $C_{12}$, $D_{1}$ and $D_{12}$ are constants. We note that these solutions can be obtained 
directly from Eqs. (\ref{frac1}), (\ref{frac2}) using a symbolic software \cite{maple}. We have chosen the parameters of the hypergeometric function in order to have an analogy with the spin-0 solution \cite{mer}. An equivalent solution with others parameters is given by Eq. (207.15) in Ref. \cite{flug}.

We choose the regular solutions at the origin $y=0$ 
\begin{equation}
\psi_{\xi}^{s}(y)=C_{1}\ y^{\nu }(1-y)^{\mu }{}\ _{2}F_{1}(\mu +\nu
+\frac{1}{2}-\frac{v_{1}}{2},\mu +\nu+\frac{1}{2}+\frac{v_{1}}{2},1+2\nu,y) , \label{sol1}
\end{equation}
\begin{equation}
\psi_{\xi}^{d}(y) =D_{1}\ y^{\nu }(1-y)^{\mu }{}\ _{2}F_{1}(\mu +\nu
+\frac{1}{2}+\frac{v_{2}}{2},\mu +\nu +\frac{1}{2}-\frac{v_{2}}{2},1+2\nu,y). \label{sol2}
\end{equation}
Then, the expressions of the components of the wave function $\psi_{\xi}$ can
be deduced as follows: from the definition (\ref{xi}) of $\psi_{\xi}^{s}(y)$ and
$\psi_{\xi}^{d}(y)$, we have 
$$\psi_{\xi}^{s}(y)+\psi_{\xi}^{d}(y)=2(\psi_{1}+\psi_{3}), \qquad
\psi_{\xi}^{s}(y)-\psi_{\xi}^{d}(y)=2(\psi_{2}+\psi_{4}),$$   
and using the Eqs. (\ref{dif1}),(\ref{dif2}) we have
\begin{equation} 
\begin{array}{l}
\psi_{1,3}(y)=\frac{1}{4} \left[1 \pm \frac{E-eV(y)}{m} \right]
[\psi_{\xi}^{s}(y)+\psi_{\xi}^{d}(y)],\\ 
\psi_{2,4}(y)=\frac{1}{4} \left[1 \pm \frac{E-eV(y)}{m}\right] 
[\psi_{\xi}^{s}(y)-\psi_{\xi}^{d}(y)], \label{comp}
\end{array}
\end{equation} 
where $\psi_{\xi}^{s}(y)$ and $\psi_{\xi}^{d}(y)$ are given by Eqs. (\ref{sol1}),(\ref{sol2}), the sign $(+)$ corresponds to first index and the sign $(-)$ to the second index.

The wave function $\psi_{\eta}$ is calculated by the same method as
$\psi_{\xi}$. We note that $\psi_{\eta}$ can be also deduced from $\psi_{\xi}$
under spatial inversion. It satisfies Eq. (\ref{sch2}) where the hamiltonian
$H_{\eta}$ is given  by Eq. (\ref{hd2}) in the one dimensional case.
Its components satisfy the four coupled differential
equations (\ref{eq1})-(\ref{eq4}) where the term
$(-ie\frac{dV(x)}{dx})$ is replaced by the term $(+ie\frac{dV(x)}{dx})$ and the components $(\psi_{1},\psi_{2},\psi_{3},\psi_{4})$ by 
$(\psi_{5},\psi_{6},\psi_{7},\psi_{8},)$, respectively. 
We give here the essential results:  
\begin{equation}
\psi_{5}-\psi_{7}=\left[\frac{E-eV(x)}{m}\right](\psi_{5}+\psi_{7}),\label{diff1}    
\end{equation}
\begin{equation}
\psi_{6}-\psi_{8}=\left[\frac{E-eV(x)}{m}\right](\psi_{6}+\psi_{8}),\label{diff2}    
\end{equation}
\begin{equation}
\left[\frac{d^{2}}{dx^{2}}+\left[ E-eV(x)\right] ^{2}-m^{2}%
-ie\frac{dV(x)}{dx}\right]\psi_{\eta}^{s}(x)=0, \label{ee3}
\end{equation}
\begin{equation}
\left[\frac{d^{2}}{dx^{2}}+\left[ E-eV(x)\right] ^{2}-m^{2}%
+ie\frac{dV(x)}{dx}\right]\psi_{\eta}^{d}(x)=0, \label{ee4}
\end{equation}
where 
\begin{equation}
\psi_{\eta}^{s}(x) = \psi _{5}+\psi_{6}+\psi_{7}+\psi_{8},\qquad  \psi_{\eta}^{d}(x)= \psi_{5}+\psi_{7}-\psi_{6}-\psi_{8} \label{eta}.
\end{equation}
We note that $\psi_{\eta}^{s}$ and $\psi_{\eta}^{d}$ satisfy the same
differential equations as $\psi_{\xi}^{d}$ and $\psi_{\xi}^{s}$, respectively. Then,
the solutions of the differential equations (\ref{ee3}) and (\ref{ee4}) are
similar to the solutions of Eqs. (\ref{e4}) and (\ref{e3}), respectively 
\begin{equation} 
\psi_{\eta}^{s}(y) =C_{2}\ y^{\nu }(1-y)^{\mu }{}\ _{2}F_{1}(\mu +\nu
+\frac{1}{2}+\frac{v_{2}}{2},\mu +\nu +\frac{1}{2}-\frac{v_{2}}{2},1+2\nu,y) , \label{sol11}
\end{equation}
\begin{equation}
\psi_{\eta}^{d}(y)=D_{2}\ y^{\nu }(1-y)^{\mu }{}\ _{2}F_{1}(\mu +\nu
+\frac{1}{2}-\frac{v_{1}}{2},\mu +\nu+\frac{1}{2}+\frac{v_{1}}{2},1+2\nu,y), \label{sol22}
\end{equation}
where $C_{2}$ and $D_{2}$ are constants.

Then, the expressions of the components of $\psi_{\eta}$ can be deduced 
as follows: from the  definition (\ref{eta}) of $\psi_{\eta}^{s}(y)$ and
$\psi_{\eta}^{d}(y)$, we have
$$\psi_{\eta}^{s}(y)+\psi_{\eta}^{d}(y)=2(\psi_{5}+\psi_{7}),\qquad
\psi_{\eta}^{s}(y)-\psi_{\eta}^{d}(y)=2(\psi_{6}+\psi_{8}),$$ 
and using Eqs. (\ref{diff1}),(\ref{diff2}) we have
\begin{equation}
\begin{array}{l}
\psi_{5,7}(y)=\frac{1}{4}\left[1 \pm \frac{E-eV(y)}{m}\right]
[\psi_{\eta}^{s}(y)+\psi_{\eta}^{d}(y)],\\ 
\psi_{6,8}(y)=\frac{1}{4} \left[1 \pm \frac{E-eV(y)}{m}\right] 
[\psi_{\eta}^{s}(y)-\psi_{\eta}^{d}(y)], \label{comp1}\\
\end{array}
\end{equation}
where $\psi_{\eta}^{s}(y)$ and $\psi_{\eta}^{d}(y)$ are given by Eqs. (\ref{sol11}),(\ref{sol22}). 
 
We study now the asymptotic behavior of the wave function when $x\rightarrow
\pm \infty .$ First, when $x\longrightarrow -\infty $ \ or $y\rightarrow 1$,
we have $(1-y)\approx \exp (x/r);$ we use the property of \ the
hypergeometric function \ which links the $y$ and $\left( 1-y\right) $
argument,
$$_{2}F_{1}(a,b,c,y)=A \ _{2}F_{1}(a,b,a+b-c+1,1-y)+B (1-y)^{c-a-b} \ _{2}F_{1}(c-a,c-b,c-a-b+1,1-y), $$
with
\begin{eqnarray*}
A =\frac{\Gamma (c)\Gamma (c-a-b)}{\Gamma (c-a)\Gamma (c-b)},\qquad
B =\frac{\Gamma (c)\Gamma (a+b-c)}{\Gamma (a)\Gamma (b)}.
\end{eqnarray*}
The corresponding constants of the waves functions $\psi_{\xi}^{s}$ and
$\psi_{\xi}^{d}$ are :
\begin{equation}
\begin{array}{l}
A_{s}=\frac{\Gamma (2\nu +1)\Gamma (-2\mu )}{\Gamma (\nu -\mu +\frac{1}{2}+\frac{v_{1}}{2})
\Gamma (\nu -\mu +\frac{1}{2}-\frac{v_{1}}{2})},\qquad 
B_{s}=\frac{\Gamma (2\nu +1)\Gamma (2\mu )}{\Gamma (\mu +\nu
+\frac{1}{2}-\frac{v_{1}}{2})\Gamma (\mu +\nu+\frac{1}{2}+\frac{v_{1}}{2} )}, \\\\
A_{d}=\frac{\Gamma (2\nu +1)\Gamma (-2\mu )}{\Gamma (\nu -\mu+\frac{1}{2}-\frac{v_{2}}{2})
\Gamma (\nu -\mu +\frac{1}{2}+\frac{v_{2}}{2})},\qquad 
B_{d}=\frac{\Gamma (2\nu +1)\Gamma (2\mu )}{\Gamma (\mu +\nu
+\frac{1}{2}+\frac{v_{2}}{2})\Gamma (\mu +\nu+\frac{1}{2}-\frac{v_{2}}{2})},
\end{array}
 \label{coef}
\end{equation}
$ \lim\limits_{y\rightarrow 1}y^{\nu }=1,$ \ $\lim\limits_{y\rightarrow
1}(1-y)^{\mu }=e^{\mu x/r},$ $\lim\limits_{y\rightarrow 1}(1-y)^{-\mu
}=e^{-\mu x/r},$ \ and $_{2}F_{1}(a,b,c,0)=1.$ \\
Thus, when $x\longrightarrow -\infty $ \ or $y\rightarrow 1,$ the waves
functions $\psi_{\xi}^{s}$ and $\psi_{\xi}^{d}$ have the following behavior:
\begin{equation}
\psi_{\xi}^{s}(x) 
\mathrel{\mathop{\longrightarrow }\limits_{x\rightarrow -\infty }}
 A_{s}e^{\mu x/r}+B_{s}e^{-\mu x/r}, \quad
\psi_{\xi}^{d}(x)
\mathrel{\mathop{\longrightarrow }\limits_{x\rightarrow -\infty }}
 A_{d}e^{\mu x/r}+B_{d}e^{-\mu x/r}.
\end{equation}
Setting $\mu =-irk_{1}$, with $k_{1}^{2}=E^{2}-m^{2}$, where $k_{1}$ is real positive
\begin{equation}
\psi_{\xi}^{s}(x)
\mathrel{\mathop{\longrightarrow }\limits_{x\rightarrow -\infty }}
A_{s}e^{-ik_{1}x}+B_{s}e^{ik_{1}x},\quad
\psi_{\xi}^{d}(x)
\mathrel{\mathop{\longrightarrow }\limits_{x\rightarrow -\infty }}
A_{d}e^{-ik_{1}x}+B_{d}e^{ik_{1}x}. \label{asy1}
\end{equation}
For the limit when $x\rightarrow +\infty ,$ or $y\rightarrow 0 ,$  $_{2}F_{1}(a,b,c,0)=1 ,$ 
\ $\lim\limits_{y\rightarrow 0}y^{\nu
}=e^{-\nu x/r},$ and  
$\lim\limits_{y\rightarrow 0}(1-y)^{\mu }=1.$
Then, the waves functions have the following behavior:
\begin{equation}
\psi_{\xi}^{s}(x)
\mathrel{\mathop{\longrightarrow }\limits_{x\rightarrow +\infty }}
e^{-\nu x/r}, \qquad
\psi_{\xi}^{d}(x)
\mathrel{\mathop{\longrightarrow }\limits_{x\rightarrow +\infty }}
e^{-\nu x/r}.
\end{equation}
Setting $\nu =-irk_{2},$ with $k_{2}^{2}=\left[ \left( E-eV_{0}\right)
^{2}-m^{2}\right] $ while $k_{2}$ is real for $E<eV_{0}-m$ or $E>eV_{0}+m$ \
and $k_{2}$ is\ imaginary for $eV_{0}-m<E<eV_{0}+m.$ Then, the waves functions
are
\begin{equation}
\psi_{\xi}^{s}(x)
\mathrel{\mathop{\longrightarrow }\limits_{x\rightarrow +\infty }}
e^{ik_{2}x}, \quad
\psi_{\xi}^{d}(x)
\mathrel{\mathop{\longrightarrow }\limits_{x\rightarrow +\infty }}
e^{ik_{2}x}. \label{d1}
\end{equation}
By the same method the asymptotic behavior of the waves functions $\psi_{\eta}^{s}$ and $\psi_{\eta}^{d}$ are
\begin{equation}
\psi_{\eta}^{s}(x)
\mathrel{\mathop{\longrightarrow }\limits_{x\rightarrow -\infty }}
 A_{d}e^{-ik_{1}x}+B_{d}e^{ik_{1}x}, \quad
\psi_{\eta}^{d}(x)
\mathrel{\mathop{\longrightarrow }\limits_{x\rightarrow -\infty }}
 A_{s}e^{-ik_{1}x}+B_{s}e^{ik_{1}x},
\end{equation}
\begin{equation}
\psi_{\eta}^{s}(x)
\mathrel{\mathop{\longrightarrow }\limits_{x\rightarrow +\infty }}
e^{ik_{2}x}, \qquad
\psi_{\eta}^{d}(x)
\mathrel{\mathop{\longrightarrow }\limits_{x\rightarrow +\infty }}
e^{ik_{2}x}. \label{d11}
\end{equation}
We note here that for the plane waves and the other wave function which are not square-integrable the renrmaliztion condition takes the form \cite{rob2} $(\psi_{E},\psi_{E'})=\pm \delta(E-E')$, while for the square-integrable wave function it takes the form $ (\psi,\psi)=\pm 1 $.

The reflection and transmission coefficients can be calculated from the
current density in the one dimensional case
\begin{equation}
j=\frac{1}{2im}\left[ \overline{\psi} O \frac{\partial \psi }{\partial x} - \frac{\partial \overline{\psi }}{\partial x}O \psi\right], \quad  A=0, \label{cour}
\end{equation}
where $ O={\mathbf 1_{2}}\otimes(\tau _{3}+i\tau _{2})\otimes {\mathbf 1_{2}}$. Using the last definition of the current and the incident wave, we find that the incident current is
\begin{eqnarray*}
j_{inc}=\frac{k_{1}}{m}\left[ B_{s}^{*} B_{d}+ B_{s}B_{d}^{*} \right]. \nonumber
\end{eqnarray*}
The reflected current is evaluated using the reflected wave  \begin{eqnarray*}
j_{ref}=-\frac{k_{1}}{m}\left[A_{s}^{*}A_{d}+ A_{s}A_{d}^{*} \right]. \nonumber
\end{eqnarray*}
Then, the reflection coefficient is
\begin{equation}
R=\frac{\left| j_{ref}\right| }{\left| j_{inc}\right| }=\frac{\left| A_{s}^{*}A_{d}+ A_{s}A_{d}^{*} \right|}
{\left| B_{s}^{*} B_{d}+ B_{s}B_{d}^{*} \right|}. \label{R}
\end{equation}
The transmission coefficient is evaluated in terms of the transmitted wave,
\begin{eqnarray*}
j_{tr}=\frac{1}{m}\left( k_{2}+k_{2}^{\dagger }\right) \exp i(k_{2}-k_{2}^{\dagger })x. \nonumber
\end{eqnarray*}
If $k_{2}$ is real,
\begin{eqnarray*}
j_{tr}=\frac{2k_{2}}{m}, \nonumber 
\end{eqnarray*}
and the transmission coefficient $T$ is
\begin{equation}
T=\frac{\left| j_{tr}\right| }{\left| j_{inc}\right| }=\frac{2k_{2}}{k_{1}}
\frac{1}
{\left| B_{s}^{*} B_{d}+ B_{s}B_{d}^{*} \right| }. \label{T}
\end{equation}
If $k_{2}$ is imaginary,
\begin{equation}
j_{tr}=0,\qquad  T=0 \qquad \mbox{ and } \quad  R=1, \label{refl}
\end{equation}
in this case we have a total reflection.

We consider now the limiting case when the smooth potential tends to step
potential, i.e., when the parameter $r$ tends to $0.$

For the first region $x<0,$ the limits of the coefficients (\ref{coef}) when $r\rightarrow 0^{+}$ are
\begin{equation}
\lim\limits_{r\rightarrow 0^{+}}A_{s}=\frac{k_{1}-k_{2}-eV_{0}}{2k_{1}}, \quad \lim\limits_{r\rightarrow
0^{+}}B_{s}= \frac{k_{1}+k_{2}+eV_{0}}{2k_{1}}, \label{lim1}
\end{equation} 
\begin{equation}
\lim\limits_{r\rightarrow
0^{+}}A_{d}=\frac{k_{1}-k_{2}+eV_{0}}{2k_{1}}, \quad  
\lim\limits_{r\rightarrow
0^{+}}B_{d}=\frac{k_{1}+k_{2}-eV_{0}}{2k_{1}}, \label{lim2}
\end{equation} 
and
the waves functions $\psi_{\xi}^{s}(x)$ and $\psi_{\xi}^{d}(x)$ are
\begin{equation}
\psi_{\xi}^{s}(x)=C_{1}\theta (-x)\left\{ \left[\frac{k_{1}-k_{2}-eV_{0}}{2k_{1}}\right]\exp
(-ik_{1}x)+\left[\frac{k_{1}+k_{2}+eV_{0}}{2k_{1}}\right]\exp (+ik_{1}x)\right\},
\label{lim11}
\end{equation}
\begin{equation}
\psi_{\xi}^{d}(x)=D_{1}\theta (-x)\left\{ \left[\frac{k_{1}-k_{2}+eV_{0}}{2k_{1}}\right]\exp
(-ik_{1}x)+\left[\frac{k_{1}+k_{2}-eV_{0}}{2k_{1}}\right]\exp (+ik_{1}x)\right\}. \label{lim22}
\end{equation}
The terms $\pm eV_{0}$ in the above relations are the contribution of the spin and if we remove them, Eqs. (\ref{lim11}),(\ref{lim22}) are reduced to the same equation as for the spin-0 case \cite{mer}.\\
For the second region $x>0,$ the waves functions are similar to Eqs. (\ref{d1})
\begin{equation}
\psi_{\xi}^{s}(x)=C_{1}\theta(x) e^{ik_{2}x},
\quad \psi_{\xi}^{d}(x)=D_{1} \theta(x) e^{ik_{2}x}.
\end{equation}
Then, the waves functions can be written in compact form for the two regions
\begin{equation}
\psi_{\xi}^{s}(x)=C_{1}\theta (-x)\left\{ \left[\frac{k_{1}-k_{2}}{2k_{1}}-\frac{eV_{0}}{2k_{1}}\right]\exp
(-ik_{1}x)+\left[\frac{k_{1}+k_{2}}{2k_{1}}+\frac{eV_{0}}{2k_{1}}\right]\exp (+ik_{1}x)\right\}
+C_{1}\theta(x) e^{ik_{2}x}, \label{step1}
\end{equation}
\begin{equation}
\psi_{\xi}^{d}(x)=D_{1}\theta (-x)\left\{ \left[\frac{k_{1}-k_{2}}{2k_{1}}+\frac{eV_{0}}{2k_{1}}\right]\exp
(-ik_{1}x)+\left[\frac{k_{1}+k_{2}}{2k_{1}}-\frac{eV_{0}}{2k_{1}}\right]\exp (+ik_{1}x)\right\}
+D_{1} \theta(x) e^{ik_{2}x}. \label{step2}
\end{equation}
Using the same method, the waves functions $\psi_{\eta}^{s}$ and
$\psi_{\eta}^{d}$ in the two regions are
\begin{equation}
\psi_{\eta}^{s}(x)=C_{2}\theta (-x)\left\{ \left[\frac{k_{1}-k_{2}}{2k_{1}}+\frac{eV_{0}}{2k_{1}}\right]\exp
(-ik_{1}x)+\left[\frac{k_{1}+k_{2}}{2k_{1}}-\frac{eV_{0}}{2k_{1}}\right]\exp (+ik_{1}x)\right\}
+C_{2}\theta(x) e^{ik_{2}x}, \label{step3}
\end{equation}
\begin{equation}
\psi_{\eta}^{d}(x)=D_{2}\theta (-x)\left\{ \left[\frac{k_{1}-k_{2}}{2k_{1}}-\frac{eV_{0}}{2k_{1}}\right]\exp
(-ik_{1}x)+\left[\frac{k_{1}+k_{2}}{2k_{1}}+\frac{eV_{0}}{2k_{1}}\right]\exp (+ik_{1}x)\right\}
+D_{2} \theta(x) e^{ik_{2}x}.  \label{step4}
\end{equation}
In the case of the step potential, we note that the presence of the delta Dirac in the differential equations (\ref{e3}),(\ref{e4}),(\ref{ee3}),(\ref{ee4}) implies that the 
functions (\ref{step1})-(\ref{step4}) are continuous at $x=0$ and 
their derivatives discontinuous.
Then, the final expression of  
the wave function $\psi(x)$ of the step potential is deduced using the expressions of the eight  
components (\ref{comp}),(\ref{comp1}) and the  
last relations (\ref{step1})-(\ref{step4}):
\begin{equation}
\psi(x)=\left(\begin{array}{c}
\psi_{1,3}(x) \\
\psi_{2,4}(x) \\
\psi_{5,7}(x) \\
\psi_{6,8}(x)
\end{array}\right)=\frac{1}{4}\left[1 \pm \frac{E-eV_{0}\theta(x)}{m}\right]
\left( \begin{array}{c}
\psi_{\xi}^{s}(y)+\psi_{\xi}^{d}(y) \\
\psi_{\xi}^{s}(y)-\psi_{\xi}^{d}(y) \\
\psi_{\eta}^{s}(y)+\psi_{\eta}^{d}(y)\\
\psi_{\eta}^{s}(y)-\psi_{\eta}^{d}(y) 
\end{array}\right) \label{final}
\end{equation}
where the sign $(+)$ corresponds to the first index and the sign $(-)$ to the second index and 
\begin{eqnarray*}
\psi_{\xi}^{s}(y) \pm \psi_{\xi}^{d}(y)&&=\theta(-x) \Bigg\{ \left[(C_{1} \pm
D_{1})\frac{k_{1}-k_{2}}{2k_{1}}-\frac{eV_{0}}{2k_{1}}(C_{1} \mp D_{1})\right]\exp
(-ik_{1}x) \\&& 
+\left[(C_{1}\pm D_{1})\frac{k_{1}+k_{2}}{2k_{1}} 
+\frac{eV_{0}}{2k_{1}}(C_{1}\mp D_{1})\right] \exp (+ik_{1}x) \Bigg\}+\theta(x)(C_{1}\pm D_{1})e^{ik_{2}x}, \\ \nonumber
\psi_{\eta}^{s}(y)\pm \psi_{\eta}^{d}(y)&&=\theta(-x) \Bigg\{ \left[(C_{2} \pm
D_{2})\frac{k_{1}-k_{2}}{2k_{1}}+\frac{eV_{0}}{2k_{1}}(C_{2}\mp D_{2})\right]\exp
(-ik_{1}x) \\ &&  
+\left[(C_{2}\pm D_{2})\frac{k_{1}+k_{2}}{2k_{1}} 
-\frac{eV_{0}}{2k_{1}}(C_{2} \mp D_{2})\right] \exp (+ik_{1}x) \Bigg\}+\theta(x)(C_{2}\pm D_{2})e^{ik_{2}x}.  \nonumber
\end{eqnarray*}
We note also that the wave function of spin-0 particle \cite{mer,mer1} can be
deduced from the wave function (\ref{final}) of spin-1/2 particle if we take only the two components
$(\psi_{1},\psi_{3})$ and remove the effect of the spin-1/2 term $(eV_{0})$. 
This analogy with spin-0 particle does not exist for the Dirac wave functions.

At the end, from Eqs. (\ref{R}),(\ref{T}) and (\ref{lim1}),(\ref{lim2}) we deduce the reflection coefficient $R$ and the transmission coefficient $T$ for the step potential \\ 
$\bullet$ For $k_{2}$ real positive $(k_{2}>0)$ and $E>eV_{0}+m,$ we have
\begin{equation}
R=\frac{(k_{1}-k_{2})^{2}-(eV_{0})^2}{(k_{1}+k_{2})^{2}-(eV_{0})^2},\qquad 
T=\frac{4k_{1}k_{2}}{(k_{1}+k_{2})^{2}-(eV_{0})^2},
\quad \mbox{ and } \quad  R+T=1. \label{ord}
\end{equation}
$\bullet$ For $k_{2}$ real negative $(k_{2}<0)$ and $m<E<eV_{0}-m,$ we have
\begin{equation}
R=\frac{(k_{1}+k_{2})^{2}-(eV_{0})^2}{(k_{1}-k_{2})^{2}-(eV_{0})^2},\qquad 
T=\frac{4k_{1}k_{2}}{(k_{1}-k_{2})^{2}-(eV_{0})^2},
\quad \mbox{ and } \quad  R-T=1,  \label{klein}
\end{equation}
which is the well known Klein Paradox.

The coefficients $T$ and $R$ given in the above relations coincide exactly with the Dirac ones \cite{flug,domb,gre1}. 
If we remove the effect of the spin-1/2 term $(eV_{0})$ from the above relations (\ref{ord}),(\ref{klein})we find the reflection and the transmission
coefficients of the spin-0 case \cite{mer}. 

Like in the spin-0 case \cite{mer}, in the following we are going to look for
boundary conditions using the Feshbach-Villars
transformation. In the case of the step potential, 
the presence of the delta Dirac in Eq. (\ref{kg}) implies that 
the KG-1/2 wave function $\Phi$ is continuous at $x=0$ and its derivative discontinuous.

The Feshbach-Villars transformation is defined for the eight components case as follows:
\begin{equation}
\psi_{\xi}(x,t)=\left(\begin{array}{c}
\psi_{1}(x,t) \\
\psi_{2}(x,t) \\
\psi_{3}(x,t) \\
\psi_{4}(x,t)
\end{array}\right)=\frac{1}{\sqrt{2}}\left(\begin{array}{c}
\varphi_{1}+\frac{i}{m}(\frac{\partial}{\partial t}+ieV)\varphi_{1} \\
\varphi_{2}+\frac{i}{m}(\frac{\partial}{\partial t}+ieV)\varphi_{2} \\
\varphi_{1}-\frac{i}{m}(\frac{\partial}{\partial t}+ieV)\varphi_{1} \\
\varphi_{2}-\frac{i}{m}(\frac{\partial}{\partial t}+ieV)\varphi_{2} 
\end{array}\right),
\end{equation}
\begin{equation}
\psi_{\eta}(x,t)=\left(\begin{array}{c}
\psi_{5}(x,t) \\
\psi_{6}(x,t) \\
\psi_{7}(x,t) \\
\psi_{8}(x,t)
\end{array}\right)=\frac{1}{\sqrt{2}}\left(\begin{array}{c}
\varphi_{3}+\frac{i}{m}(\frac{\partial}{\partial t}+ieV)\varphi_{3} \\
\varphi_{4}+\frac{i}{m}(\frac{\partial}{\partial t}+ieV)\varphi_{4} \\
\varphi_{3}-\frac{i}{m}(\frac{\partial}{\partial t}+ieV)\varphi_{3} \\
\varphi_{4}-\frac{i}{m}(\frac{\partial}{\partial t}+ieV)\varphi_{4} 
\end{array}\right),
\end{equation}
which can be written in the abridged two-component form 
\begin{equation}
\left(\begin{array}{c}
\psi_{j}\\
\psi_{j+2}
\end{array}\right)=\frac{1}{\sqrt{2}}\left(\begin{array}{c}
\varphi_{k}+\frac{i}{m}(\frac{\partial}{\partial t}+ieV)\varphi_{k} \\
\varphi_{k}-\frac{i}{m}(\frac{\partial}{\partial t}+ieV)\varphi_{k} 
\end{array}\right),  \label{two}
\end{equation}
where the couple $(j,k)$ takes the following values:
$(j,k)=\{(1,1),(2,2),(5,3),(6,4)\}$, respectively, $(\psi_{j},\psi_{j+2})^T$ are the
eight components of the wave function $\psi$ with $j=1,2,5,6$ and 
$\varphi_{k}=\varphi_{k}(x,t)$ are the four components of the wave function
$\Phi$ with $k=1,2,3,4$.

From the last definition (\ref{two}) it follows that
\begin{equation}
\varphi_{k} =\frac{1}{\sqrt{2}}(\psi _{j}+\psi _{j+2}), \label{phi1}
\end{equation}
\begin{equation}
(i\frac{\partial }{\partial t}-eV)\varphi_{k} =\frac{m}{\sqrt{2}}\left(\psi
_{j}-\psi_{j+2}\right) .
\end{equation}
The stationary KG-1/2 wave function $\Phi$ has the
form $\Phi (x,t)=e^{-iEt}\Phi (x)$ for each component and the last equation is
written as
\begin{equation}
(E-eV)\varphi_{k} =\frac{m}{\sqrt{2}}\left( \psi _{j}-\psi _{j+2}\right). \label{phi2}  
\end{equation}
The continuity at $x=0$  of the component $\varphi_{k}$ of the KG-1/2 wave function 
defined in Eq. (\ref{phi1}) gives
\begin{equation}
\psi _{j}(0^{+})+\psi _{j+2}(0^{+})=\psi _{j}(0^{-})+\psi_{j+2}(0^{-}). \label{sum1}
\end{equation} The continuity at $x=0$  of the component $\varphi_{k}$ of the KG-1/2 wave function
defined in Eq. (\ref{phi2}) gives
\begin{equation}
\psi _{j}(0^{+})-\psi _{j+2}(0^{+})=\frac{E-eV_{0}}{E} \left[\psi
_{j}(0^{-})-\psi _{j+2}(0^{-})\right]. \label{sum3}
\end{equation} 
From Eqs. (\ref{sum1}) and (\ref{sum3}) we can write the boundary
conditions in the matrix form
\begin{equation}
\left( 
\begin{array}{c}
\psi _{j}(0^{+})\\
\psi _{j+2}(0^{+})
\end{array}
\right)=
\left(
\begin{array}{cc}
 1-\frac{eV_{0}}{2E} & \frac{eV_{0}}{2E} \\
 \frac{eV_{0}}{2E} & 1-\frac{eV_{0}}{2E} \end{array}
\right)
\left(
\begin{array}{c}
\psi _{j}(0^{-}) \\
\psi _{j+2}(0^{-})
\end{array}
\right), \label{bound1}
\end{equation}
where $j=1,2,5,6$.

For the following particular cases the boundary conditions are more simple:\\
$\bullet$ For $E=\frac{eV_{0}}{2}$: 
$$\psi _{j}(0^{+})=\psi _{j+2}(0^{-}),\qquad \psi _{j+2}(0^{+})=\psi _{j}(0^{-}).$$ 

This case will be discussed at the end.\\
$\bullet$ For $E=eV_{0}$:
$$\psi _{j}(0^{+})=\psi _{j+2}(0^{+})=\frac{1}{2}[\psi _{j}(0^{-})+\psi _{j+2}(0^{-})].$$

The boundary conditions (\ref{bound1}) are between the components of the wave function
$\psi$. It can be written for all the eight components as follows: 
let us introduce the following notations
$$\psi_{I}=\sum_{j=1,2,5,6}\psi_{j}=\psi_{1}+\psi_{2}+\psi_{5}+\psi_{6},$$
$$\psi_{II}=\sum_{j=1,2,5,6}\psi_{j+2}=\psi_{3}+\psi_{4}+\psi_{7}+\psi_{8},$$
from relation (\ref{sum1}), the sum of the eight
components of the wave function satisfies  
\begin{equation}
\psi_{I}(0^{+})+\psi_{II}(0^{+})=\psi_{I}(0^{-})+\psi_{II}(0^{-}). \label{ss1}
\end{equation}
From relation (\ref{sum3}), the difference
between $\psi_{I}$ and $\psi_{II}$ satisfies   
\begin{equation}
\psi_{I}(0^{+})-\psi_{II}(0^{+})=
\frac{E-eV_{0}}{E}\left[\psi_{I}(0^{-})-\psi_{II}(0^{-})\right]. \label{ss3}
\end{equation}
From relations (\ref{ss1}) and (\ref{ss3}) we can also write the above boundary
conditions in the matrix form
\begin{equation}
\left(
\begin{array}{c}
\psi_{I}(0^{+})\\
\psi_{II}(0^{+})
\end{array} \right)=\left(
\begin{array}{cc} 
1-\frac{eV_{0}}{2E} & \frac{eV_{0}}{2E} \\ 
\frac{eV_{0}}{2E} & 1-\frac{eV_{0}}{2E}
\end{array} \right) \left(
\begin{array}{c}
\psi_{I}(0^{-}) \\ 
\psi_{II}(0^{-})
\end{array} \right). \label{SSI}
\end{equation}
Using the boundary conditions (\ref{bound1}), we find that 
the charge density $\rho $ is discontinuous
\begin{equation}
\rho (0^{+})=\frac{E-eV_{0}}{E} \rho (0^{-}), \label{charge}
\end{equation}
it gives the charge sign of the transmitted particle as a 
function of the energy $E$, the potential $V_{0}$, the charge $e$ and the charge density of the incident particle. 
We note that the same relation (\ref{charge}) can be also obtained for spin-0 particle. 
In the case of the Dirac equation, the density is continuous at $x=0$. Then, the Dirac density is a probability 
density and cannot be a charge density .

At the end, we verify that the wave functions (\ref{final}) of the step potential 
satisfy the boundary conditions (\ref{bound1}),(\ref{SSI}),(\ref{charge}) and the current(\ref{cour}) is continuous at $x=0$.
    
The boundary conditions (\ref{ss1}), (\ref{ss3}) can be also interpreted by analogy with
electromagnetic waves (when they traverse two different regions) as follows:
the sum of the components of the wave function  $\psi _{s}=\psi_{I}+\psi _{II}$ 
is continuous like the tangential
component of the electric field  but the difference of 
the two components $\psi_{d}=\psi _{I}-\psi _{II}$ 
is discontinuous like the normal component of the magnetic field. 

Now we discuss the boundary condition (\ref{charge}) of the charge density 
and compare its predictions to those obtained from the transmission and reflection coefficients for 
the three following cases(we assume for all cases \cite{domb} that $eV_{0}>2m$  and  $E>m$ $\quad$ i.e.  $k_{1}>0$):\\
 
1. \quad $E> eV_{0}+m $ \\
$k_{2}$ is real and $ 0<\frac{E-eV_{0}}{E}<1$, from Eq. (\ref{charge}) the charges densities $\rho (0^{+})$ and $\rho (0^{-})$ have the same sign, i.e. the charge of the transmitted particle has the same sign as the charge of the incident one. 
For $k_{2}>0$, the transmission and the reflection coefficients are given by relation (\ref{ord}). For $E>>eV_{0}+m$: 
$\rho (0^{+}) \approx \rho (0^{-})$, $T \approx 1$ and $R \approx 0$. The two methods give the same results.\\

2. \quad $eV_{0}-m<E<eV_{0}+m$ \\
$k_{2}$ is imaginary and from Eq. (\ref{charge}) we consider two cases:\\ 
$\bullet$ For $eV_{0}<E<eV_{0}+m$: we have $\frac{E-eV_{0}}{E}>0$; the charges densities $\rho (0^{+})$ and $\rho (0^{-})$ have the same sign.\\ 
$\bullet$ For $eV_{0}-m<E<eV_{0}$: we have $\frac{E-eV_{0}}{E}<0$; the charge densities $\rho (0^{+})$ and $\rho (0^{-})$ have opposite sign, 
this means that we have creation of particle-antiparticle pairs near the step barrier if the potential is strong enough.

On the other hand, from Eq. (\ref{refl}) we have a total reflection $R=1$ and $T=0$ and the wave function (\ref{final}) is decreasing (evanescent wave) in the second region \cite{gre1,ni}. The only case for which we have a total reflection from Eq. (\ref{charge}) is for $E=eV_{0}$:  $\rho (0^{+})=0$. \\

3. \quad $m<E<eV_{0}-m$ \\
$k_{2}$ is real and $\frac{E-eV_{0}}{E}<0$, from Eq. (\ref{charge})  
the charges densities $\rho (0^{+})$ and $\rho (0^{-})$ have opposite sign. This means that we have also creation of pairs near the step barrier. For $k_{2}<0$, the transmission and the reflection coefficients  are given by Eqs. (\ref{klein}). Then, the creation of pairs in the Klein Paradox \cite{gre1} is proved  
from the boundary condition of the charge density (\ref{charge}). We note that Guang-jiong Ni et al  \cite{ni,ni1} have discussed the Klein paradox, for the spin-0 case, using the current and the charge density in the two regions but 
they haven't used the boundary condition of the charge density.

Let us now study an interesting particular case for the spin-0 particle\cite{mer}:\\
$\bullet$ For the particular value $E=\frac{eV_{0}}{2}$ from the interval $m<E<eV_{0}-m$ (we have assumed that $eV_{0}>2m$): we have
$k_{2}=\pm k_{1}$, $T=1$ and $R=0$ i.e. the incident particle is transmitted to the second region (called the resonance transmission) but
from relation (\ref{charge}) we have $\rho (0^{+}) =-\rho (0^{-})$: 
this means that we have creation of pairs near the step barrier
and the transmitted particle is the antiparticle of the incident one. 
The result of the case $k_{2}=-k_{1}$ need an interpretaion.\\ 

 In summary, in order to find the wave functions of the step potential
without the use of boundary conditions, we introduce the smooth potential as
an intermediate stage. Then, we solve the one dimensional Feshbach-Villars
equation for spin-1/2 particle subjected to the smooth potential. The eight-component 
wave function is given in terms of the hypergeometric functions. In
the limiting case $r\longrightarrow 0,$ the wave functions of the step
potential are deduced in each region. The transmission and reflection coefficients   
are identical to the Dirac ones. We have also an analogy between  
the wave functions and the transmission and reflection coefficients of the spin-1/2
and the spin-0 particles.
Boundary conditions relative to the
step potential are extracted using the Feshbach-Villars
transformation and the continuity of the KG-1/2 wave
function at $x=0$. The main result is that boundary
conditions for the step potential are: \\
$\bullet $ the sum of the eight components $\psi_{s}=\psi _{I}+\psi _{II}$ is continuous:
$$\psi_{s}(0^{+})=\psi_{s}(0^{-}),$$
$\bullet $ the difference of the two components $\psi_{d}=\psi_{I}-\psi_{II}$ is discontinuous:
$$\psi_{d}(0^{+})=\frac{E-eV_{0}}{E}\psi_{d}(0^{-}),$$  
$\bullet$ the charge density $\rho$ is discontinuous:
$$\rho (0^{+})=\frac{E-eV_{0}}{E} \rho (0^{-}),$$ \\
and for $\frac{E-eV_{0}}{E}<0$ we have creation of particle-antiparticle pairs if the potential is strong enough. 
Then, the number of particles becomes variable and this implies that we must use quantum field theory \cite{domb1,domb,gre2}.

At the end, we note that we have omitted the singular solution in order to make comparison with the Dirac results. We propose to study the contributions of this solution separately to this paper.\\

{\bf\large Acknowledgments}\\ 
One of the authors (A.B.) wishes to thank Prof. A. Di Giacomo for the hospitality at the physics department of the University of Pisa where this work was done 
and the Italian foreign ministry for financial support. He also thanks Dr. B.A. Robson for the papers \cite{rob3,rob4} and Dr. D.S. Staudte for the preprints \cite{stau1,stau2}. The authors thank the referee for his remarks.

\end{document}